\documentclass[conference]{IEEEtran}
\IEEEoverridecommandlockouts
\usepackage{cite}
\usepackage{amsmath,amssymb,amsfonts}
\usepackage{algorithmic}
\usepackage{graphicx}
\usepackage{textcomp}
\usepackage{hyperref}
\usepackage{xcolor}
\def\BibTeX{{\rm B\kern-.05em{\sc i\kern-.025em b}\kern-.08em
    T\kern-.1667em\lower.7ex\hbox{E}\kern-.125emX}}
\begin{document}

\newcommand{\sys}{\textsc{Spark}}

\title{\sys{}: Real-Time Monitoring of Multi-Faceted Programming Exercises}

\author{
\IEEEauthorblockN{Yinuo Yang$^{*}$\thanks{$^{*}$Work done as an undergraduate student at the University of Michigan.}}
\IEEEauthorblockA{
\textit{College of Engineering} \\
\textit{University of Notre Dame}\\
Notre Dame, IN USA \\
yinooyang@nd.edu}

\and
\IEEEauthorblockN{Ashley Ge Zhang}
\IEEEauthorblockA{\textit{School of Information} \\
\textit{University of Michigan}\\
Ann Arbor, MI USA \\
gezh@umich.edu}
\and
\IEEEauthorblockN{Steve Oney}
\IEEEauthorblockA{\textit{School of Information} \\
\textit{University of Michigan}\\
Ann Arbor, MI USA  \\
soney@umich.edu}
\and
\IEEEauthorblockN{April Yi Wang}
\IEEEauthorblockA{\textit{Department of Computer Science} \\
\textit{ETH Zurich}\\
Zurich, Switzerland\\
april.wang@inf.ethz.ch}
}
\maketitle

\begin{abstract}
Monitoring in-class programming exercises can help instructors identify struggling students and common challenges. However, understanding students' progress can be prohibitively difficult, particularly for multi-faceted problems that include multiple steps with complex interdependencies, have no predictable completion order, or involve evaluation criteria that are difficult to summarize across many students (e.g., exercises building interactive web-based user interfaces). We introduce \sys{}, a coding exercise monitoring dashboard designed to address these challenges. \sys{} allows instructors to flexibly group substeps into checkpoints based on exercise requirements, suggests automated tests for these checkpoints, and generates visualizations to track progress across steps. \sys{} also allows instructors to inspect intermediate outputs, providing deeper insights into solution variations. We also construct a dataset of 40-minute keystroke coding data from N=22 learners solving two web programming exercises and provide empirical insights into the perceived usefulness of \sys{} through a within-subjects evaluation with 16 programming instructors.
\end{abstract}

\begin{IEEEkeywords}
programming education
\end{IEEEkeywords}

\section{Introduction}
Programming instructors often use \emph{in-class exercises}---short hands-on coding tasks conducted during class time---to actively engage students and reinforce the concepts being taught~\cite{de2003characteristics, hmelo2004problem, michael2006s, felder2009active, prince2004does, peng2025glitteraiassistedplatformmaterialgrounded}.
However, ensuring students gain meaningful learning outcomes from these exercises is not easy, given the variability in coding abilities, paces, and problem solving approaches~\cite{hmelo2004problem}.
This variation can make it challenging for instructors to provide timely and personalized feedback. 
Without such assistance, students may struggle to develop essential metacognitive skills, such as formulating effective problem solving strategies, tracking their progress, and assessing whether goals have been met~\cite{hmelo2004problem, jonassen2015all, schon2015teaching, simon1973structure}.
This can lead to frustration and a potential loss of confidence in their abilities~\cite{IntelligentCoaching}. 
Therefore, it is essential for instructors to effectively monitor students' progress and promptly recognize the difficulties they encounter.

However, successfully monitoring students can be challenging, particularly for problems that are \emph{multi-faceted}.
We refer to ``multi-faceted'' problems as those involving non-sequential workflow paths with nested substeps---where some steps are interdependent, others independent, and where uniform evaluation criteria cannot be easily applied across many students. 
For example, in a web programming exercise, students may need to: (1) create the HTML layout, (2) add CSS for styling, and (3) implement JavaScript for interactivity.
Adding CSS may involve edits to both the CSS and HTML files and depends on the layout being complete, but is independent of the JavaScript.
Students can choose their own order and often alternate between tasks as they work.

Prior work has emphasized the importance of real-time monitoring tools, but existing solutions struggle to effectively summarize student progress for complex, multi-faceted programming problems.
Code clustering tools (e.g., \cite{glassman2015overcode, zhang2024cflow}) and progress visualizations (e.g., \cite{vizprog}) can summarize many code samples  but do not give instructors control over which aspects to group by or summarize.
Techniques for monitoring code in real-time (e.g., \cite{codeopticon,wang2021puzzleme}) give instructors real-time feedback but do not summarize students' progress and can be overwhelming in large classes.
Further, most prior work does not address additional difficulties of monitoring in-class exercises.
Implementations for features often span multiple files or modules~\cite{welling2003php, grinberg2018flask} but most prior work is focused on short, one-file snippets~\cite{vizprog,glassman2015overcode,codeopticon}.
Further, instructors should be able to \emph{explore} variations in students' code output and intermediate states to gain deeper insights into students' approaches and challenges.

To address these challenges, we introduce \sys{}, a coding exercise monitoring dashboard designed for multi-faceted programming exercises. \sys{} enables instructors to customize multi-level checkpoints with testing code suggestions, allowing them to track student progress for individual tasks. 
At each checkpoint, the testing code evaluates the intended outcomes, ensuring that students meet the specific objectives of the exercise. 
A progress visualization diagram that summarizes students' progress across tasks is generated using evaluation data from the testing code. 
Additionally, \sys{} allows instructors to customize inspections of intermediate variables and outputs, providing active engagement and deeper insights into students' program state.

While \sys{} is adaptable to various types of multi-faceted programming exercises, its implementation is specifically tailored for web programming. 
It includes features such as customized real-time inspections of output variations, as well as viewing DOM attributes and clustered previews of selected elements for evaluating web element performance across different students. 
Additionally, it allows for the simulation of element interactions within the testing code before conducting inspections, accommodating the event-driven nature of web programming tasks~\cite{alimadadi2014understanding, burg2013interactive, hibschman2019isopleth}. 

   We created a dataset consisting of 22 students' keystroke data for two web programming problems in a 40-minute session. Using this data, we simulated a real-time classroom setting and conducted a within-subject user study with 16 participants to evaluate \sys{}'s effectiveness in helping instructors monitor students' programming progress. 
We found that \sys{} helps participants 1) identify students' challenges more accurately and 2) feel more confident in their understanding of students' programming progress. Participants also reported that \sys{} provides more detailed information and valuable, customizable insights into variations in students' code states. 
This work can help instructors improve real-time teaching by deepening their understanding of students' mental models and encouraging active engagement in the monitoring process. 
This work makes the following contributions:
\begin{itemize}
\item A pipeline that uses customized checkpoints with nested steps to visualize student progress and inspect immediate output variations in real-time. 
\item \sys{}, a system based on this pipeline, designed to monitor student progress for web programming exercises.
\item A dataset containing coding keystroke data from 22 students for two web programming exercises.
\item A within-subject user study involving 16 participants validating the effectiveness of \sys{}.
\end{itemize}

\section{Related Work}
\subsection{Understanding Students' Programming Progress}

Prior research has introduced various methods to visualize students’ programming progress and support real-time monitoring, clustering, and runtime inspection. Many tools used 2D maps to track code changes and similarities~\cite{taniguchi2022visualizing, huang2013syntactic}, and tools like VizProg~\cite{vizprog} helped track different programming approaches. While these maps reflect the relative proximity of code states based on edit distance, they do not convey absolute positioning or capture the nuances of non-linear workflows in multi-faceted programming tasks. Such tasks involve interdependent subgoals that students often tackle out of sequence, making traditional linear visualizations insufficient. To address this, \sys{} introduces a checkpoint-based framework for tracking progress across varied sequences.

Real-time monitoring is essential for timely feedback and maintaining student engagement~\cite{de2003characteristics, hmelo2004problem, michael2006s, felder2009active, prince2004does, IntelligentCoaching}. Tools like Codeopticon~\cite{codeopticon}, RIMES~\cite{rimes}, and VizProg~\cite{vizprog} offer dashboards for live observation, yet they often present fixed data views. PuzzleMe~\cite{wang2021puzzleme} provides insights via peer-generated test cases, but it is designed for peer support, not instructor control. These systems fall short when instructors need to customize what they monitor, leading to either data overload or insufficient detail. \sys{} overcomes this by letting instructors selectively view runtime outputs and variables, tailoring the feedback to immediate teaching needs.

To manage growing class sizes, many studies have employed clustering to summarize student solutions. Techniques such as AST edit distances~\cite{huang2013syntactic}, Overcode~\cite{glassman2015overcode}, and CFlow~\cite{zhang2024cflow} group similar code to surface common patterns. Others, like~\cite{kaleeswaran2016semi, head2017writing, nguyen2014codewebs}, focus on functional patterns or errors. However, these methods often assume structural similarity, which is not always present in open-ended, multi-section exercises like web development. \sys{} supplements clustering with test case results, enabling instructors to evaluate functional correctness regardless of divergent implementation paths.

Finally, understanding the runtime behavior of students' code is critical, especially when students struggle to articulate their issues~\cite{markel2021inside}. Systems like Callisto~\cite{wang2020callisto} link questions to code, but real-time classroom scenarios demand more scalable solutions. Visualizing runtime values helps comprehension~\cite{storey1999cognitive}, yet with many students, it becomes hard to decide where to focus~\cite{miller1956magical, jackson2012theory}. Tools like RunEx~\cite{runex} and TeachNow~\cite{teachnow} provide scalable inspection and assistance. \sys{} integrates workflow visualization with variable inspection, allowing instructors to first identify students with unusual progress patterns, and then drill down into variable-level details, offering a guided path from overview to diagnosis.

\subsection{Runtime Variable Visualization}

Variable visualizations play a crucial role in code comprehension, as inspecting variable states is essential to understanding how a program behaves~\cite{kernighan1999practice}. Prior works have visualized variable values adjacent to code~\cite{lerner2020projection, guo2021ten, guo2013online, jiang2023log}, while tools like Omnicode~\cite{kang2017omnicode} and Theseus~\cite{lieber2014addressing} present runtime behavior through scatterplot matrices or inline displays. CrossCode~\cite{hayatpur2023crosscode} extends this by visualizing multi-level execution traces. However, in classroom settings, instructors face significant cognitive load when trying to inspect runtime variables for every student~\cite{miller1956magical}. To address this, \sys{} enables clustered runtime value visualizations, helping instructors see variable states, outputs, and program behavior at scale. In event-driven, interactive web programming tasks, visual outputs are often key to understanding runtime behavior. Tools like Colaroid~\cite{wang2023colaroid}, CoCapture~\cite{cocapture}, and InterState~\cite{oney2014interstate} demonstrate the power of visual representations in supporting comprehension and communication. \sys{} integrates this approach, allowing instructors to view students’ rendered output directly, improving their understanding of dynamic interface behavior.

In the context of AI-generated code, research highlights the importance of building trust through runtime feedback~\cite{wang2024investigating}, with visualizing intermediate values shown to help validate AI outputs~\cite{ferdowsi2024validating}.
Beyond AI, comparisons of runtime states also aid in understanding and debugging student or unfamiliar code. For instance, DITL helps data scientists compare datasets~\cite{wang2022diff}, and Doppio visualizes changes in UI flows~\cite{chi2018doppio}. These findings support the value of runtime comparison. In \sys{}, instructors can monitor runtime behavior as students write code, and when creating test cases, they can combine AI-generated suggestions with reference validation to verify both intermediate and final states—offering a reliable path to ensure code correctness.
\section{System Design}
\subsection{System Design Goals}

\sys{} aims to support instructors in effectively monitoring students' progress during multi-faceted programming exercises in real time. Its design goals stem from the reflective analysis of instructional challenges around the difficulty of tracking diverse learning paths~\cite{vizprog, GilPermanentMonitoring}. 

\subsubsection{DG1: Customizable Structured Progress Monitoring}

Effectively monitoring individual student progress during programming exercises is critical for classroom management and student success~\cite{hmelo2004problem, GilPermanentMonitoring, wang2021puzzleme}, but it remains challenging.
First, the diversity of teaching contexts requires flexible monitoring approaches~\cite{deed2020teacher}, as classroom needs and instructional goals vary~\cite{koedinger2012knowledge, schwartz2016abcs}. This highlights the limitations of one-size-fits-all solutions and underscores the need for customizable tools.
Second, interpreting student progress data can impose significant cognitive load on instructors~\cite{codeopticon}, particularly in exercises involving complex workflows and varied learning paths.
Even with real-time data, making sense of it remains difficult. Tools like Glancee~\cite{ma2022glancee} and VizProg~\cite{vizprog} help visualize progress to reduce cognitive demands, but Lee et al.\cite{LeeMonitoring} emphasize that flexibility and customizability are essential to avoid information overload. This supports Dillenbourg’s\cite{dillenbourg2013design} argument that loosely structured activities are hard to manage without checkpoints.

These insights motivate DG1: \emph{customizable structured progress monitoring}, calling for tools that provide customizable, organized insights tailored to the teaching contexts.

\subsubsection{DG2: Gain a Holistic Understanding of Class Progress}
While individual code submissions offer detailed insights, instructors need efficient tools to detect broader patterns that indicate conceptual misunderstandings across the class~\cite{card1999readings}. This need arises in two key contexts: in real-time teaching, instructors must balance individual support with class-wide awareness~\cite{LeeMonitoring, kazemitabaar2024codeaid}; in lab sessions, they often rely on large scale of retrospective reviews such as recordings or edit histories due to limited real-time visibility~\cite{riese2018teaching}.

These challenges underscore the need for tools that aggregate and simplify synchronous and asynchronous programming data to reveal meaningful patterns while minimizing information loss~\cite{taniguchi2022visualizing, huang2013syntactic, vizprog}. Prior work demonstrates this principle: Taniguchi et al.\cite{taniguchi2022visualizing} and Huang et al.\cite{huang2013syntactic} used 2D maps to visualize code evolution and similarity, and Zhang et al.\cite{vizprog} showed the value of tracking student progress at multiple granularities for classroom management. These align with visualization principles aimed at simplifying complex data without sacrificing essential information\cite{demvsar2006data, fayyad2002information, costagliola2008monitoring}, and with Tissenbaum’s call for real-time visualization to support instructional orchestration~\cite{tissenbaum2016real}.

\

From these insights, we derive DG2: \emph{aggregated progress tracking}, enabling instructors to efficiently detect class-wide patterns and bottlenecks.
\subsubsection{DG3: Query Multiple Properties of Students' Code}

DG3 builds on research in program comprehension~\cite{ferdowsi2024validating, chi2018doppio, lerner2020projection, guo2021ten, guo2013online, jiang2023log, kang2017omnicode, lieber2014addressing} and instructional needs in classroom settings~\cite{koedinger2012knowledge, miller1956magical, schwartz2016abcs}. Studies show that understanding program behavior requires more than reviewing source code or final outputs—it involves examining intermediate runtime states~\cite{ferdowsi2024validating, chi2018doppio}. To support this, prior work has introduced techniques such as displaying variable values next to code~\cite{lerner2020projection, guo2021ten, guo2013online, jiang2023log} and embedding runtime visualizations in code editors~\cite{kang2017omnicode, lieber2014addressing}.

Instructors face additional challenges in classrooms, where they must assess knowledge mastery~\cite{koedinger2012knowledge, schwartz2016abcs}, manage limited time and cognitive resources~\cite{miller1956magical}, and monitor many students simultaneously. 
Real-time visualization of intermediate code behavior can support this process, enhancing teaching efficiency.

Thus, DG3: \emph{code querying capabilities} promotes tools that go beyond passive code review, allowing instructors to actively query checkpoint correctness, runtime state, and output. This empowers them to apply their expertise in diagnosing and supporting student learning.

\subsection{Overview of \sys{}}
\begin{figure*}[t]
    \centering
    \includegraphics[width=1\linewidth]{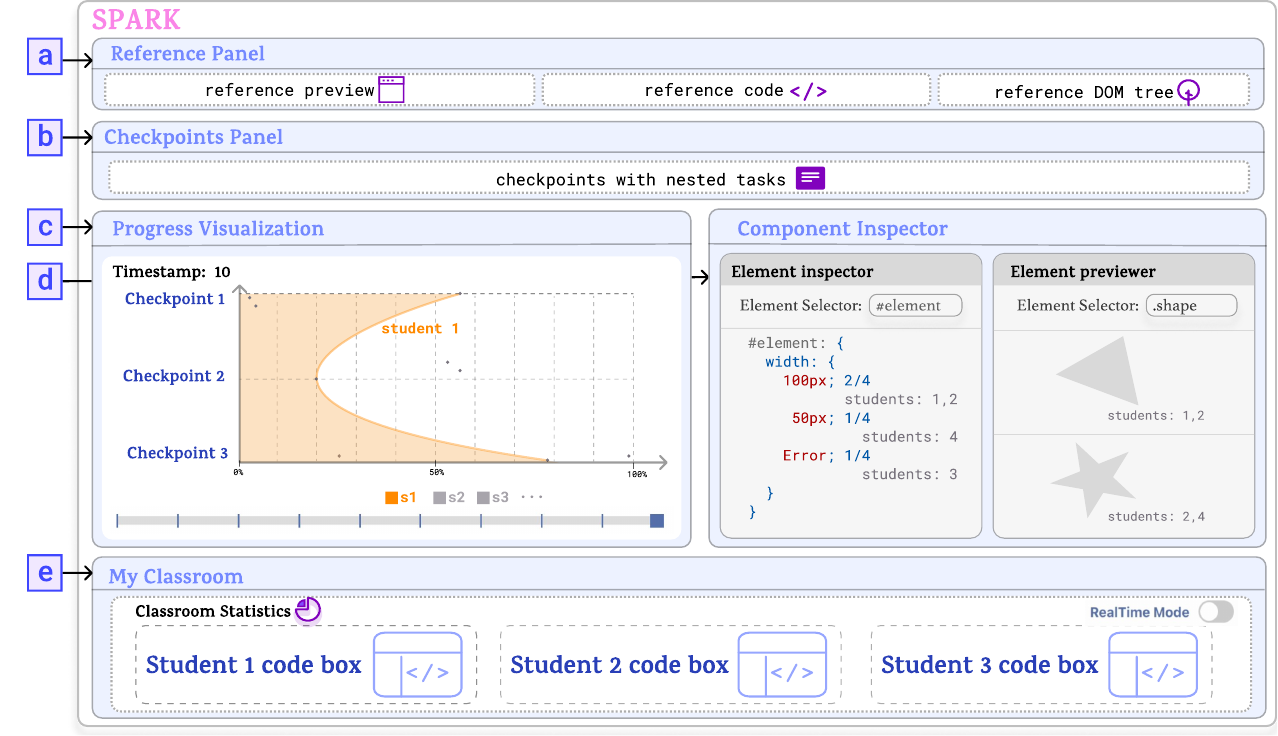}
    \caption{\sys{} consists of five panels. Here's an overview of the \sys{} dashboard: the \textit{Reference Panel} (a), which provides instructors with the expected code answer, the webpage, and a DOM tree preview for the programming exercise; the \textit{Checkpoints Panel} (b), which allows for the creation and display of nested-task checkpoints; the \textit{Progress Visualization} (c), which presents a visualization of students' programming progress; the \textit{Component Inspector} (d), which enables instructors to customize inspections of students' output variations; and the \textit{My Classroom} (e), which contains student code and statistics on overall task performance.
    }
    \label{overview}
\end{figure*}
Informed by our design goals, we developed \sys{} as a dashboard to help instructors monitor students' programming progress in real-time for multi-faceted programming problems,  with its implementation focus on web programming exercises.
\sys{} consists of five panels:
\begin{itemize}
   \item \textbf{Reference} panel (Fig.\ref{overview}.a), where instructors can enter their reference code answers to \textit{Reference Code} board. The related preview (in \textit{Reference Page Preview} board) and DOM tree (in \textit{DOM Tree Reference} board) would be automatically generated.
    \item \textbf{Checkpoints} panel (Fig.\ref{overview}.b), which allows the creation and display of nested-task checkpoints. Each task includes a description and testing code for assessment.
   \item \textbf{Progress Visualization} panel (Fig.\ref{overview}.c), which features a progress visualization diagram.
   \item \textbf{Components Inspector} panel (Fig.\ref{overview}.d), which allows instructors to inspect students' output variations.
   \item \textbf{My Classroom} panel (Fig.\ref{overview}.e), which contains classroom statistics and student code boxes.
\end{itemize}

To illustrate the experience of using \sys{}, we describe how a hypothetical instructor, Emily, conducts a multi-faceted programming exercise in class in real-time. Emily wishes to monitor students' programming progress, understand their programming progress, and provide timely assistance. For instance, she wants to identify common issues students are facing, as well as those who are falling behind, and offer support accordingly. 
Below, we describe how \sys{} can help Emily monitor the classroom, highlighting both its features and implementation. 
In the scenario below, descriptions of \sys{}'s key features are integrated with screenshots and implementation details for each feature. 
Only instructors can see the features of \sys{}.

\subsection{Creating Checkpoints with Nested Tasks}
\begin{figure*}[t]
    \centering
    \includegraphics[width=1\linewidth]{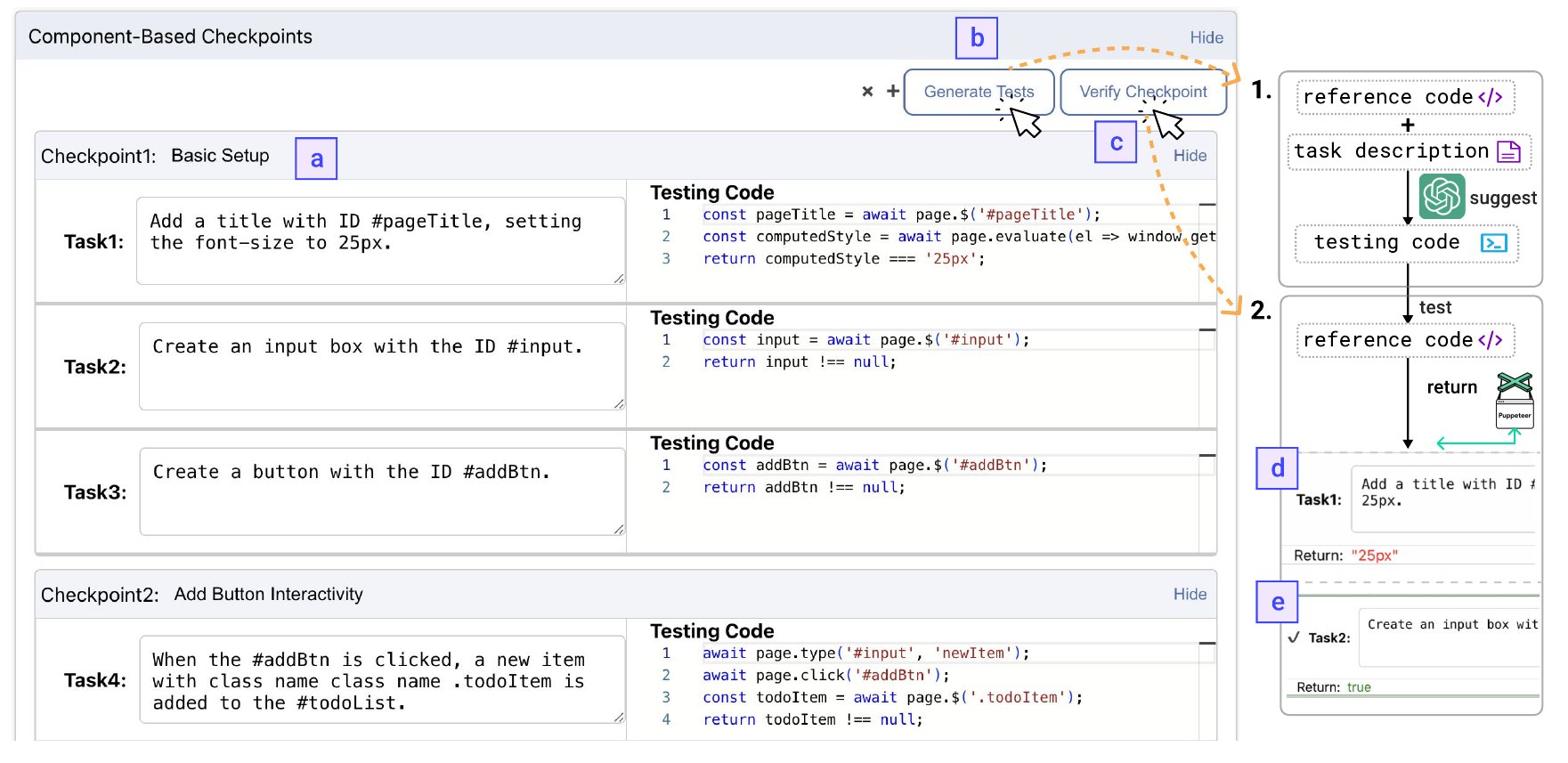}
    \caption{\textit{Checkpoints} panel. In the \textit{Checkpoints} panel, instructors can freely create nested-task checkpoints. Each checkpoint consists of multiple tasks, and each task (a) is made up of two parts: \textit{Task Description} and \textit{Testing Code}. Instructors can click the \textit{Generate Tests} (b) button to view AI-suggested testing code, which can be manually modified. They can then click the \textit{Verify Checkpoint} (c) button to test the reference code to see if it passes (e) or fails (d).
    }
    \label{checkpoints}
\end{figure*}
Before the class begins, to use \sys{} to monitor students' programming progress, Emily first creates checkpoints with grouped steps in the \textit{Checkpoints} panel (Fig.\ref{checkpoints}) (DG1). The process of creating checkpoints with nested tasks involves three steps. First, Emily inputs the reference code answers into the \textit{Reference Code} board in the \textit{Reference} panel, enters the task description into the task box (Fig.\ref{checkpoints}.a), and clicks the \textit{Generate Test} button (Fig.\ref{checkpoints}.b) to use AI for generating testing code based on the task description and the reference code. Next, she could review the AI-generated testing code to ensure it meets her expectations, making any necessary modifications. Finally, to verify the accuracy of the testing code, Emily clicks Verify Checkpoint (Fig.\ref{checkpoints}.c) to check if the testing code successfully passes against the reference code, displaying a success message (Fig.\ref{checkpoints}.e). If the test fails, the system allows instructors to retrieve information through the return statement (Fig.\ref{checkpoints}.d), which aids in debugging and identifying the issue. Throughout the editing and verification process, Emily could continually use the \textit{Reference Page Preview}, \textit{Reference Code}, and \textit{DOM Tree Reference} boards to preview the expected programming exercise outcome, assisting in the creation and verification of checkpoints.

The testing code serves two primary functions: first, it evaluates whether a student's code meets the step requirements by assessing the behavior of specific elements; second, it simulates interactions before performing the evaluation. Additionally, the testing code is used in element inspections, as will be explained in Section~\ref{inspector_section}, allowing instructors to observe output variations with the required interactions simulated.

\textit{Implementation:} \sys{} uses the OpenAI API~\cite{OpenAIChatGPT2024} to provide testing code suggestions\footnote{More details could be found in the supplementary material: \href{https://inonnno.github.io/Academic-project-page-Spark/static/pdfs/VLHCC_SPRAK_Supplementary.pdf}{link}.}
 and employs Puppeteer~\cite{Puppeteer2024} to simulate and evaluate code execution. The \textit{Reference Page Preview} is implemented using an iframe, and the \textit{DOM Tree Reference} is generated based on the \textit{Reference Code}.

\subsection{Real-time Monitoring of Students' Progress} 
\begin{figure*}[t]
    \centering
    \includegraphics[width=1\linewidth]{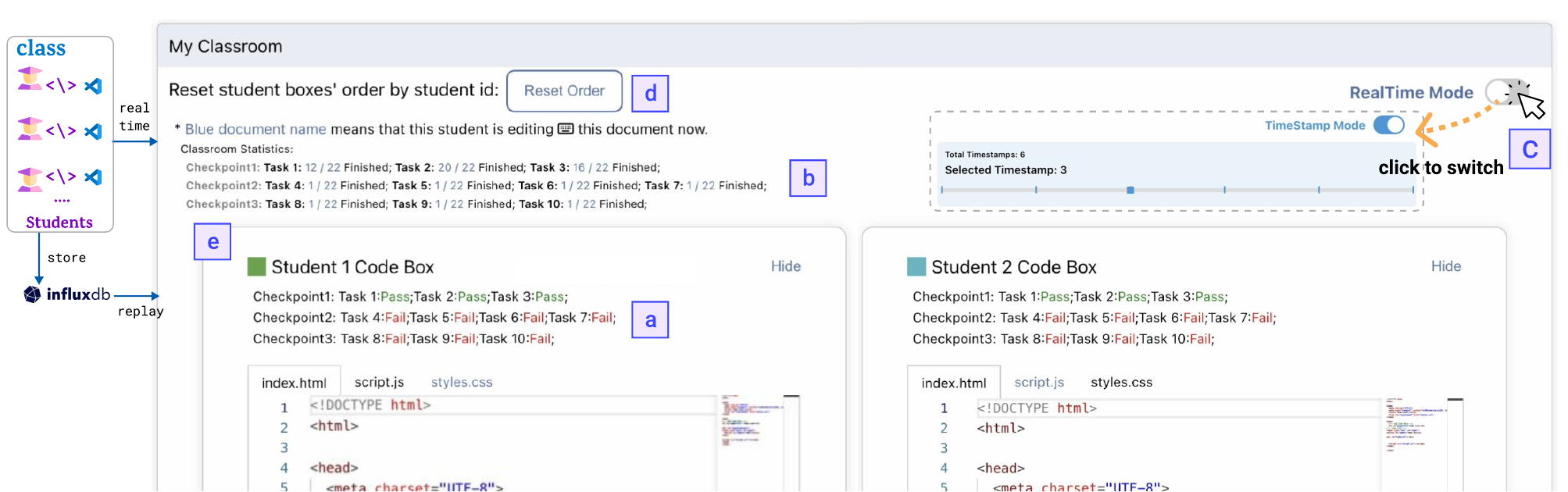}
    \caption{\textit{My Classroom} panel. Each student working on the programming problem has a \textit{Student Code Box} (e), which includes the student's code and \textit{task completion status} (a) indicating whether they have passed the task. There is also a \textit{Classroom Statistics} board (b) showing the class’s performance across different tasks.
The \textit{My Classroom} panel can switch between two modes (c): \textit{Real-Time Mode}, where instructors can view students' code in real-time, and \textit{Timestamp Mode}, which records students' code history every minute. In \textit{Timestamp Mode}, Instructors can use the slider to review the students' code history. If instructors change the order of the student code boxes in the \textit{Progress Viz} panel or \textit{Components Inspector} panel, they can use the \textit{Reset Order} button (d) to revert the Code Boxes to their default order.}
    \label{classroom}
\end{figure*}

Once Emily creates the checkpoints, she simply shares a folder with setup files and starter code. When students open it in VS Code with the required extension, \sys{} begins receiving real-time programming data. This data is reorganized and displayed in code boxes within the \textit{My Classroom} panel (Fig.\ref{classroom}), similar to Codeopticon\cite{codeopticon}.

Each code box (Fig.\ref{classroom}.e) shows a student’s live code and task completion status (Fig.\ref{classroom}.a) across checkpoints. The \textit{Classroom Statistics} board (Fig.\ref{classroom}.b) offers a high-level view of task progress (DG2). The panel supports two modes: real-time (live keystroke updates) and timestamp (minute-by-minute snapshots), allowing Emily to review students’ code history via a slider (Fig.\ref{classroom}.c). A blue-highlighted file name indicates the currently active file.

By default, code boxes are ordered by student ID. Emily can rearrange them in the \textit{Progress Visualization} or \textit{Components Inspector} panels, and reset the layout via the \textit{Reset Order} button (Fig.\ref{classroom}.d).
While students work, Emily monitors progress using the \textit{Progress Visualization} panel (auto-updated every minute) and the \textit{Components Inspector} panel for more detailed inspection (Figs.\ref{viz}, \ref{ins}).

\textit{Implementation:} \sys{} uses a custom VS Code extension to capture and transmit keystroke-level data (edit content, location, and timestamp). Only this lightweight edit data is sent to \sys{}, which organizes and displays it in real time within the \textit{My Classroom} panel (Fig.\ref{classroom}).

\subsection{Progress Visualization View}
\label{pro_viz}
\begin{figure*}[t]
    \centering
    \includegraphics[width=1\linewidth]{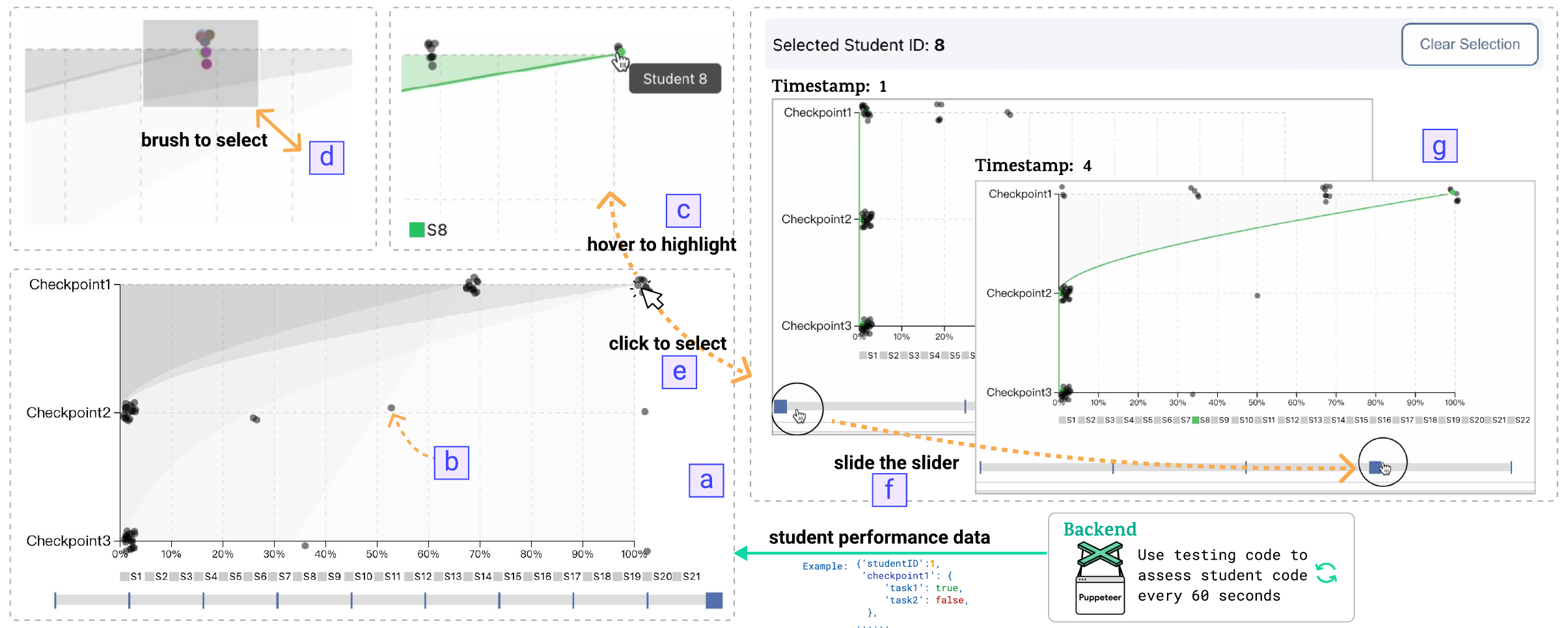}
    \caption{\textit{Progress Visualization} panel. The \textit{Progress Visualization} panel includes a diagram (a) that displays students' progress across checkpoints. Each dot along the checkpoint line represents a student's task completion rate for that checkpoint at a particular timestamp (b). Instructors can hover over a dot to highlight a student (c), use the brush tool to select multiple students within that area (d), and click to select an individual student (e), with the selected student remaining highlighted. Additionally, instructors can adjust the slider to view the visualization diagrams at different timestamps (f).
    }
    \label{viz}
\end{figure*}

In the \textit{Progress Visualization} panel (Fig.\ref{viz}), each student's progress within a checkpoint is represented by a dot (Fig.\ref{viz}.a), placed left to right from 0\% to 100\% task completion (Fig.\ref{viz}.b). A student’s overall progress is visualized as a shaded area; overlapping areas indicate similar progress levels, with darker shades showing higher student density (DG2).

Hovering over a dot highlights the student's trajectory line across checkpoints and changes their shaded area to a unique color while hiding others (Fig.\ref{viz}.c) (DG1). Emily can also use the brush tool to select and view groups of students (Fig.\ref{viz}.d) (DG2) or use the slider (Fig.\ref{viz}.f) to explore progress over time.

To investigate a specific student who struggled with checkpoint 2, Emily can click their dot or label to lock the highlight (Fig.\ref{viz}.e), then use the slider to trace their progress over time (Fig.\ref{viz}.g). The student’s code box is also brought to the top for direct inspection.

\textit{Implementation:} \sys{} uses Puppeteer~\cite{Puppeteer2024} to run predefined test cases on students’ code, using the results to generate visualizations of checkpoint completion.

\subsection{Component Inspector View}
\label{inspector_section}
\begin{figure*}[t]
    \centering
    \includegraphics[width=0.9\linewidth]{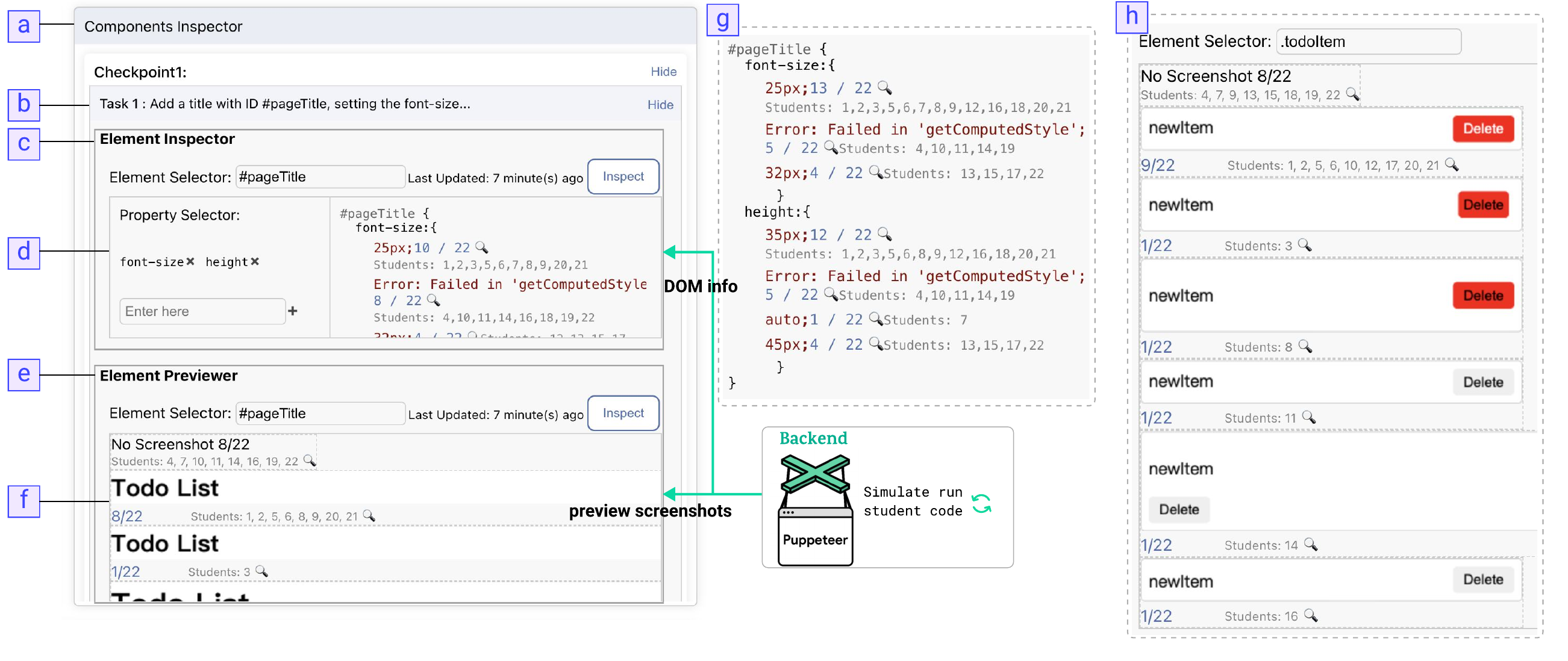}
    \caption{\textit{Components Inspector} Panel. The \textit{Components Inspector} (a) is structured according to checkpoints. Each task is linked to a corresponding inspector board (b) that includes two key features: the \textit{Element Inspector} (c) and the \textit{Element Previewer} (e). 
Instructors can view the inspection results in (d) (a full view of the results is shown in (g)) and preview the element in (f).
The inspector automatically simulates interactions before performing the inspection. For example, (h) displays clustered screenshots of the newly added \texttt{.todoitem} following the interaction \textit{"Add new item after clicking the add button (Task 4)"}. Screenshots of identical elements are grouped together to facilitate easier analysis.
    }
    \label{ins}
\end{figure*}

With a general view of student progress, Emily turns to the \textit{Components Inspector} panel (Fig.\ref{ins}.a) for deeper analysis of specific elements. This panel offers two customizable inspection features: DOM property inspection (Fig.\ref{ins}.d) and visual previews (Fig.\ref{ins}.f), which can be used separately or together.

Emily selects the relevant task (Fig.\ref{ins}.b), enters the element selector, and uses the \textit{Element Inspector} to check property variations via the \textit{Property Selector} (Fig.\ref{ins}.d, g), with matching student counts shown (DG3). Alternatively, she can preview how elements render across students using the \textit{Element Previewer} (Fig.\ref{ins}.e, f). Clicking \textit{Inspect} runs the analysis, and she can use the magnifier button to bring students with matching issues to the top of the \textit{My Classroom} panel for closer review.

The inspector panel mirrors the checkpoint structure, with each task linked to an inspector board. Results reflect simulated interactions from the checkpoint’s testing code (e.g., Fig.\ref{ins}.h shows interaction with the ``add todo item" element).

\textit{Implementation:} \sys{} sends students’ code to Puppeteer~\cite{Puppeteer2024}, which simulates test interactions before inspection. For the \textit{Element Previewer}, visual similarities are clustered using Resemble.js~\cite{resemblejs}.

\subsection{Recording and Replaying}
\label{replay}
\sys{} supports both real-time monitoring and replay via keystroke recording. The replay feature helps address key challenges: the cognitive load of real-time tracking, unequal attention to students, and oversight during TA-led sessions.
For example, when instructors like Emily step away to assist students, they may miss critical moments. With \sys{}, they can review class data afterward or merge asynchronous session recordings (DG2), enabling retrospective analysis to identify common struggles and provide targeted support. This ensures no student progress is overlooked.

\textit{Implementation:}
\sys{} logs keystroke data to a database (e.g., InfluxDB). In replay mode, it retrieves and chronologically replays this time-series data for simulation.

\subsection{Example Usage Contexts}
\sys{} supports both in-person and online classrooms, including asynchronous settings like MOOCs, enabling real-time progress monitoring and learning support. This flexibility is especially valuable for reserved students who may hesitate to seek help~\cite{socialdisconnect, picciano2002beyond, ni2013comparing}, allowing instructors to track and assist learners regardless of when or how they engage.

\sys{} can also scale beyond intermediate web programming by adjusting checkpoint granularity—for example, supporting finer-grained steps in beginner machine learning tasks like building a digit recognition pipeline.
With shared rubrics and replay features, \sys{} promotes consistent evaluation and coordinated instruction across teaching teams.

\section{Evaluation}
\subsection{Dataset of Real-Time Programming Data}\label{dataset}

To simulate a real-time classroom environment for web programming exercises, we conducted a data collection session prior to the user study~\cite{vizcode}. We recruited 22 students
with programming experience, including 10 beginners, 11 intermediate, and 1 advanced in web programming.

Sessions were held via Zoom, with each participant completing two same 20-minute introductory web programming tasks.
 A research team member collected data using a customized VS Code extension. Participants could use resources like Google and Copilot\cite{GitHubCopilot2024}, but not LLMs to generate code, balancing realistic usage with data reliability.

Keystroke-level data—averaging 810 keystrokes per student per task—was stored in InfluxDB with timestamps, edit locations, and anonymized IDs. \sys{} then replayed this data chronologically to simulate real-time progress. This dataset offers a fine-grained view of coding behavior, addressing the lack of real-time detail in traditional datasets and providing a valuable benchmark for future research on programming learning and problem-solving.

\subsection{User Study}

To evaluate \sys{}’s effectiveness, we conducted a within-subjects study with 16 participants experienced in teaching and web programming. Using the dataset from Section~\ref{dataset}, participants simulated real-time classroom monitoring by observing replays and answering quiz questions about student progress and challenges.
\subsubsection{Recruitment}

We recruited 16 participants (9F, 6M) from Computer Science and Information Science departments based on their teaching background and web programming experience. Participants included instructors, tutors, TAs, and experienced graduate students. Fifteen had prior teaching experience, with 1 to 6+ years of web programming experience.

\subsubsection{Study Protocol}
The user study used a within-subjects design with three sessions. In S1 and S2, participants used the Baseline and \sys{} systems to monitor student progress. And in S3, participants evaluated the preparation phase by creating checkpoints and test code using \sys{}.
\begin{itemize}
    \item \textbf{Baseline:} A simplified version of \sys{} with \textit{Reference} and \textit{My Classroom} panels; real-time and timestamped code views only, without performance metrics.
        \item \textbf{\sys{}:} Full version of \sys{} with all the features.
\end{itemize}



Tasks were consistent across S1 and S2, with the order counterbalanced. 
S3 was always used as the final activity. 
Each session included a 5-minute tutorial. Four participants joined in person, while 12 participated remotely.

In S1 and S2, participants observed replays of 22 students' progress and answered quiz questions during and after the session. The quiz questions included nine questions combining fact-based queries, diagnostic tasks, and open-ended reflections.
Each session included with a questionnaire (5-point Likert scale) and a brief interview. S1 and S2 were limited to 25 minutes for comparability. 
In S3, after a tutorial, participants created a checkpoint and generated corresponding test code, then completed a usability questionnaire.

\subsubsection{Data Collection and Analysis} 

 

\begin{table*}[htbp]
\centering
\caption{Mixed-Effects Linear Regression Model Results for Tool Effect (\sys{} vs. Baseline)}
\label{tab:tool_effect}
\begin{tabular}{lcccccc}
\hline
\textbf{Statement} & \textbf{$b$} & \textbf{SE} & \textbf{$z$} & \textbf{$p$} & \textbf{95\% CI Low} & \textbf{95\% CI High} \\
\hline
Monitor all students' programming progress comprehensively.
    & 2.44  & 0.31  & 7.94  & $<$.001 & 1.84  & 3.04  \\
Gain detailed insights into students' programming trajectories.
    & 2.00  & 0.29  & 6.80  & $<$.001 & 1.42  & 2.58  \\
Identify students who are falling behind.
    & 2.06  & 0.33  & 6.35  & $<$.001 & 1.43  & 2.70  \\
Identify challenges faced by students.
    & 2.06  & 0.35  & 5.87  & $<$.001 & 1.37  & 2.75  \\
Assess the element functionality among students.
    & 2.75  & 0.27  & 10.37 & $<$.001 & 2.23  & 3.27  \\
Inspect variations in element behavior output among students.
    & 3.00  & 0.27  & 10.93 & $<$.001 & 2.46  & 3.54  \\
Number of correct answers in the quiz (5 in total)
  & 2.19  & 0.26  & 8.35  & $<$.001 & 1.67  & 2.70  \\
Number of unsure answers in the quiz (5 in total)
   & -2.25 & 0.38  & -5.93 & $<$.001 & -2.99 & -1.51 \\
\hline
\end{tabular}
\end{table*}

During screening, we collected participants’ teaching and web programming experience. In each session, a researcher took observation notes and graded quiz responses. S1 and S2 used nearly identical quizzes tailored to their respective exercises, each containing 5 multiple-choice and 4 open-ended questions. We recorded quiz accuracy and time spent per question using screen recordings.

\begin{figure*}[t]
    \centering
    \includegraphics[width=0.9\linewidth]{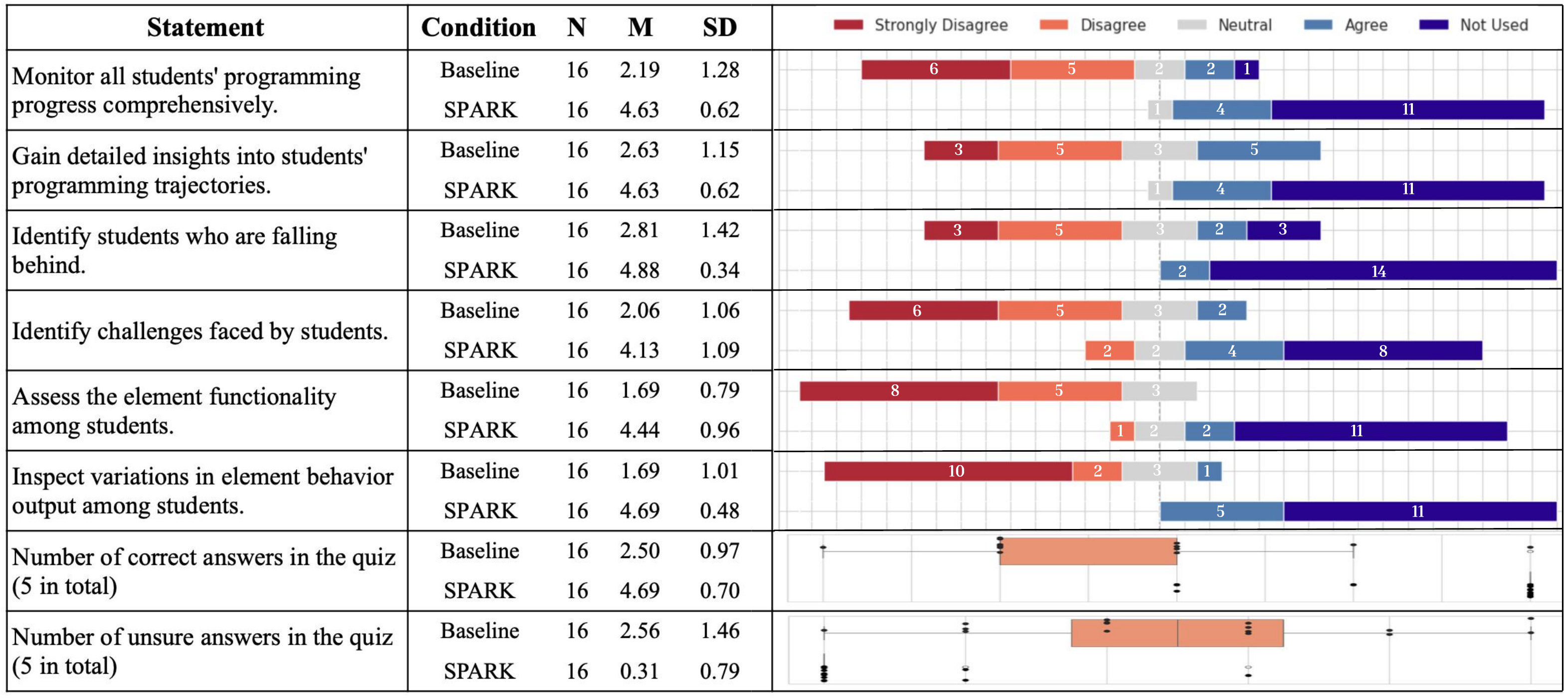}
    \caption{Perceptions of the Baseline and \sys{} system. 
    Participant rated on a 5-point scale.
    (M: mean. SD: standard deviation).
    }
    \label{sta}
\end{figure*}

Data analysis included questionnaire ratings, quiz performance, and self-reported confidence. 
A mixed-effects linear regression model revealed significant effects of system type (\sys{} vs. Baseline). The order in which tools were used also had a significant effect in most cases, with lower ratings observed when SPARK was used first---likely due to comparison effects. In contrast, problem type showed no consistent influence. Full results are shown in Table~\ref{tab:tool_effect} and Fig.~\ref{sta}.


We also conducted a thematic analysis of 16 semi-structured interviews. Transcripts were coded and iteratively clustered into themes to extract key insights.
\subsection{Results}

\subsubsection{\sys{} helps instructors identify students' challenges more accurately}
In the quiz, participants assessed students' performance, identified issues, and recognized shared challenges. 
We counted the number of correct answers participants provided in each session, and there was a significant difference between \sys{} and Baseline (p $<$ .001), with \sys{} showing a marked improvement in answer accuracy. 

We observed that in the Baseline session, when asked to assess students' programming performance, 11 out of 16 participants only glanced at the top half of the student code box, making observations based on this limited view. In contrast, 15 out of 16 participants in the \sys{} session used the \textit{Component Inspector} to gain insights into students' code behavior. Several participants noted that \sys{} has a learning curve, but once they became familiar with the system and the checkpoints, it significantly improved the quality of monitoring, making the effort worthwhile (P1, P5, P9-P10). P9 mentioned that \textit{``..., it is kind of hard (to learn)...but it gives good overview of students' states."}.

\subsubsection{\sys{} improves instructors' confidence in understanding students' programming progress}
We found that participants gained significantly more confidence in the monitoring results with \sys{}, as shown in Fig.\ref{sta}. There was a notable difference in the number of unsure answers in the quiz between \sys{} and the Baseline system (p $<$ .001). In the Baseline system, many participants made incorrect assessments of students' performance (P1, P4-6, P11-14), often misinterpreting syntax they believed to be correct but that was actually incorrect. Additionally, some participants (P4, P12) were confident in their observations and assessments while using the Baseline system, yet still made inaccurate assessments.


When comparing \sys{} to the Baseline system, many participants highlighted the value of test cases for evaluating specific tasks. As P8 noted, \textit{``Using test cases to show inter-student progress...that's a really good idea."} P10 added, \textit{``They provide a more intimate understanding...showing real-time progress and highlighting issues."}

In post-session interviews (S1 and S2), all 16 participants expressed interest in using \sys{} in real classrooms, while only 7 were open to using the Baseline system—2 of whom would do so only in small classes. As P12 commented, \textit{``It's better than nothing...but with many students, I won’t have time."} P7 remarked, \textit{``Looking at too many students’ code is exhausting. I’d rather see nothing."}

\subsection{System Usability and Study Insights}
\begin{figure*}[t]
    \centering
    \includegraphics[width=0.8\linewidth]{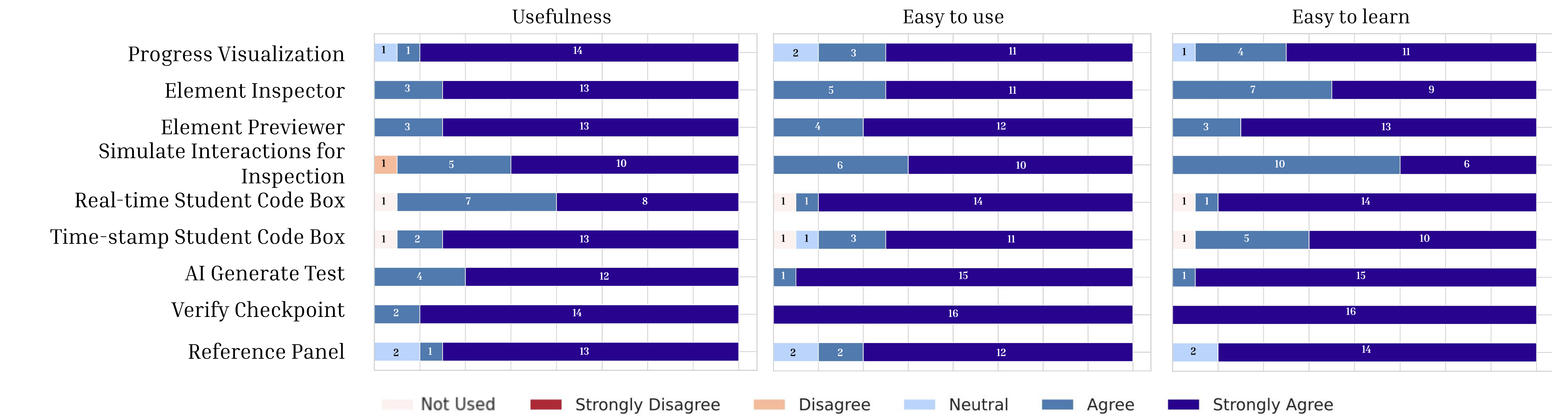}
    \caption{Results from the questionnaire of the Likert-scale responses to “usefulness”, “easy to use”, and “easy to learn” after each session.
    }
    \label{fig7}
\end{figure*}

\subsubsection{\sys{} enables more detailed programming progress monitoring}
In the post-session questionnaire, 15 of 16 participants agreed that \sys{} provided detailed insights into student progress. Compared to the Baseline, \sys{} enabled faster, more accurate checkpoint assessments—14 of 16 answered correctly in a quiz using \sys{}, versus 7 with the Baseline. Baseline users also took twice as long on average.

Participants (P2–6, P11–15) praised the \textit{Progress Visualization} for its clarity and intuitive interactions (e.g., hover and brush), with P2 noting, \textit{``Grouping tasks into checkpoints saves unnecessary effort reviewing similar code."}

All participants used the visualization to identify struggling students and reviewed code histories via the timestamp feature. Several (e.g., P13, P16) noted students followed varied workflow paths to reach correct solutions.

\sys{} offered deeper insights beyond task progress. P6 pointed out that abrupt code trajectory changes could indicate copying. In an open-ended quiz, 12 participants used \sys{} to decide which topics to revisit, and 9 used classroom statistics to identify difficult concepts. P2 noted, \textit{``Seeing how long students struggled helps decide what to emphasize next time."}

\subsubsection{\sys{} enables a better understanding of variations among students' code output}

In the post-session questionnaire, participants rated \sys{} significantly higher than the Baseline system for inspecting variations in students' output (p $<$ .001). Many found the \textit{Component Inspector} helpful for understanding performance, with P6 and P8 noting it \textit{``provides a direct way to understand students' performance"}, and P2 highlighting its value for visualizing layout and properties.

The checkpoint structure also reduced cognitive load and made tracking progress more intuitive. 
P15 noted, \textit{``Organizing tasks into checkpoints is intuitive and allows for more detailed insights."} While Baseline users struggled to recall common issues, 15 of 16 participants using \sys{} successfully identified at least one issue faced by over half the class.

\subsubsection{\sys{} enables instructors to active engage with the monitoring process}
During the \sys{} condition, participants used the \textit{Components Inspector} an average of 3.4 times during the 20-minute session. Many appreciated its customization, with P2 noting, \textit{``It’s great that I could inspect only one element—much easier to compare."} 
In exploring issues in \textit{checkpoint 2}, 15 of 16 participants used the inspector, most selecting multiple task boards. As P12 observed, \textit{``...the add button interactivity is more difficult, so I’d take a look at this."} 
\sys{}’s flexible inspection tools enabled participants to focus on specific problem areas, boosting their confidence in identifying issues and offering targeted feedback.

\subsubsection{\sys{} makes creating step-nested checkpoints as well as their test cases easy}

In the third session (S3), participants used \sys{} to create a checkpoint with one task. Features were highly rated for usefulness, ease of use, and ease of learning (Fig.\ref{fig7}). Most participants strongly agreed that \textit{AI Generate Test} (12/16), \textit{Verify Checkpoint} (14/16), and \textit{Reference Panel} (13/16) supported easy test case creation and clarified assessment goals. 

Participants found the AI-generated tests \textit{``super convenient and time-saving (P10, P11)"}, while the verification and reference features gave them \textit{``confidence [they] could use these in real classrooms (P13)"}. All 16 participants expressed willingness to use these features for creating step-nested checkpoints.

\textit{User Challenges and Feedback.} Participants identified several challenges when using \sys{}. First, they raised concerns about the scalability of the scatter plot, which became increasingly cluttered and difficult to interpret as the number of students grew. Second, some participants noted that the system’s rich features and modular interface, while powerful, occasionally introduced additional visual and cognitive load—particularly during real-time monitoring. These observations suggest the need for systems that can better support large-scale classrooms while maintaining usability and minimizing cognitive effort.

\section{Discussion}

\subsection{Visualizing Multi-facet Programming at Scale}
The expansion of programming education has led to larger class sizes, making it difficult for instructors to monitor student progress—especially in multi-faceted tasks with loosely structured workflows, such as data analysis pipelines, GUI applications, or hardware programming. These tasks involve complex codebases and exploratory coding behaviors that traditional clustering techniques struggle to capture. For instance, in data analysis, students may experiment with features that don’t impact final outputs, which runtime- or AST-based clustering may overlook or misrepresent.

Our study shows that checkpoint-based visualizations provide a more effective solution. 
By allowing instructors to define key stages of the task, they can track how students iteratively approach each checkpoint over time. This design accommodates non-linear workflows and highlights meaningful progress. Future work may explore expanding checkpoint mechanisms using AI-assisted techniques~\cite{ferdowsi2024validating} or peer assessment~\cite{wang2021puzzleme} to provide richer insights into student performance.

Our user study confirms that \sys{} performs well in small-to-medium-sized classrooms. In testing with 22 students, a single Puppeteer server maintained sub-30-second computation times. These benchmarks suggest the system can scale further for low-latency, real-time tracking in larger classrooms.

\subsection{Visualizing Student Code with Spatial Meanings}
\begin{figure}
    \centering
    \includegraphics[width=1.0\linewidth]{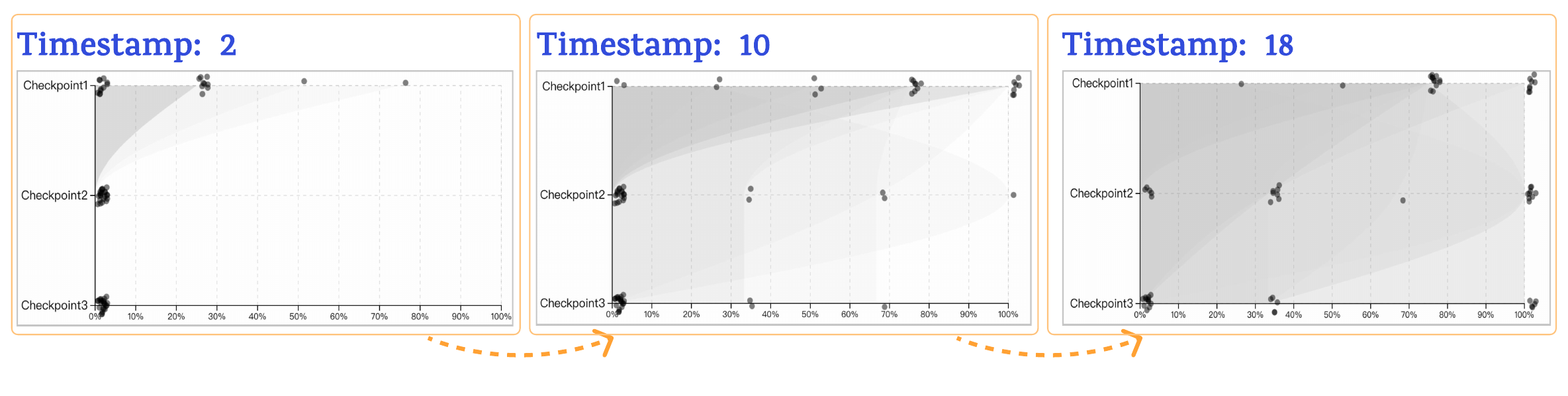}
    \caption{
How the shaded areas in the visualization diagram change over time in the \textit{Image Carousel} example.
    }
    \label{shade}
\end{figure}
By visualizing test case results across checkpoints, \sys{} provides a reliable way to track student progress in multi-faceted programming exercises. This approach reveals students' workflow sequences and clarifies task interdependencies, enabling instructors to quickly assess progress across different code sections. For example, in the \textit{Image Carousel} exercise (Fig.\ref{shade}), some students completed \textit{Checkpoint 2} first, while most followed the expected sequence. At \textit{timestamp 18}, performance was highest on \textit{Checkpoint 2} and lowest on \textit{Checkpoint 3}. The spatial layout helps instructors understand both overall progress and specific challenges at each stage.

While VizProg~\cite{vizprog} uses absolute code positions for progress, \sys{} aligns progress with individual workflow sequences and using shaded spatial regions to signal differences in progress. This enhances instructors’ situational awareness and offers a more intuitive understanding of student behavior.
Although this method scales well to large classes, it may lead to information overload in very large cohorts. 
Future work could explore adaptive filtering and summary views to surface key trends and outliers.

\subsection{Engagement with Instructors}
Our findings show that participants actively engaged with \sys{}'s interactive features, finding them effective for managing information and retrieving key details. Unlike prior systems~\cite{codeopticon, glassman2015overcode, vizprog} that support mostly passive monitoring, \sys{} enables a hands-on approach, allowing instructors to create custom checkpoints and inspect student work in-session—reducing cognitive load and boosting confidence in instructional decisions.

As AI tools become more common, it's essential to balance automation with human oversight. While AI can assist with predictions or evaluations, meaningful instruction relies on active engagement. Effective learning analytics should go beyond correctness metrics to offer insights into student reasoning and strategies. \sys{} supports this by combining AI-generated test suggestions with instructor-led exploration.

Future work could incorporate features like automated feedback~\cite{kazemitabaar2024codeaid}, predictive analytics, and inactivity tracking, while ensuring instructors remain central to the monitoring process.

\subsection{Limitations}

A current limitation of \sys{} is the preparation effort. While it is currently tailored for intermediate web programming, future work could explore AI-assisted preparation workflows to balance customization with efficiency, and extend \sys{} to advanced domains such as machine learning.

The user study also has two main limitations. First, the initial sessions focused on using pre-generated checkpoints and test code, with only the third session evaluating their creation. This offers limited insight into how instructors might generate and use custom tests in authentic teaching contexts. Future studies could examine instructor interactions when designing their own tests to better understand \sys{}'s support for real-time monitoring. Second, since \sys{} can inform teaching pace and concept review, future research could explore its long-term classroom use to understand how instructors adapt and integrate its features into everyday teaching.
\section{Conclusion}
This paper presents a real-time visualization approach for multi-faceted programming exercises in classroom settings. We developed \sys{}, a dashboard that lets instructors define checkpoints, suggest automated tests, and visualize student progress across varied workflow sequences. \sys{} also supports inspection of intermediate outputs, offering deeper insight into students’ code states. Our evaluation shows that \sys{} facilitates easy checkpoint creation, detailed progress monitoring, and active, customizable engagement. By visualizing progress and variables, \sys{} helps instructors better understand students’ mental models, reduce cognitive load, and deliver more effective, personalized feedback at scale.

\section*{Acknowledgment}
We thank Xinyi Chen for the valuable feedback on the \sys{} project. We are also grateful to all study participants and collaborators for their time and contributions.

\bibliographystyle{IEEEtran}
\bibliography{ref}
\end{document}